\documentclass[showpacs,twocolumn,floats,superscriptaddress]{revtex4}
\usepackage{graphicx}
\usepackage{amsmath}
\usepackage{bm}

\newcommand{\bq}{\begin{equation}}
\newcommand{\ee}{\end{equation}}
\newcommand{\fr}[2]{\frac{#1}{#2}}
\newcommand{\eps}{\varepsilon}

\renewcommand{\vec}[1]{\mathbf{#1}}

\begin{document}
\title{Functionalized Graphene in Quantizing Magnetic Field: The case of
bunched impurities }
\date{\today }

\author{P. G. Silvestrov}

\affiliation{Institute for Mathematical Physics, TU Braunschweig,
38106 Braunschweig, Germany}

\affiliation{Physics Department and Dahlem Center for Complex
Quantum Systems, Freie Universit\"{a}t Berlin,
14195 Berlin, Germany}


\begin{abstract}
Resonant scattering at the atomic absorbates in graphene was
investigated recently in relation with the transport and gap
opening problems. Attaching an impurity atom to graphene is
believed to lead to the creation of unusual zero energy localized
electron states. This paper aims to describe the behavior of the
localized impurity-induced levels in graphene in a quantizing
magnetic field. It is shown that in the magnetic field the
impurity level effectively hybridizes with one of the n=0 Landau
level states and splits into two opposite-energy states. The new
hybridized state is doubly occupied, forming a spin-singlet and
reducing the polarization of a Quantum Hall ferromagnet in undoped
graphene. Taking into account the electron-electron interaction
changes radically the spectrum of the electrons surrounding the
impurity, which should be seen experimentally. While existing
publications investigate graphene uniformly covered by adatoms,
here we address a possibly even more experimentally relevant case
of the clusterized impurity distribution. The limit of a dense
bunch of the impurity atoms is considered, and it is shown, how
such a bunch changes the spectrum and spin polarization of a large
dense electron droplet surrounding it. The droplet is encircled by
an edge state carrying a persistent current.

\end{abstract}
 \maketitle

\section{Introduction}\label{Section_Introduction}

First theoretical works following the discovery of graphene almost
a decade ago~\cite{Novosel05, Zhang05} were concentrated on the
relativistic character of its electronic spectrum. It was however
quickly realized that this two-dimensional material can offer a
plethora of interesting effects, going far beyond the
quasirelativistic behavior of the bulk electrons~\cite{review09}.
One class of such effects comes from the investigation of the
graphene edges, where for example in the case of zigzag edge one
finds a band of dispersion-less zero energy edge states.
Remarkably, as was shown already in the early
work~\cite{Pereira06}, even the shortest possible edge in
graphene, which is the closed edge of the hole created by removing
a single carbon atom, is sufficient to create a single localized
zero energy electron state with the algebraic wave function
$\psi\sim 1/(x+iy)$. The existence of such localized low energy
states with a power law coordinate dependance of the density is an
indication of the resonant scattering at the Dirac point in
graphene.

A number of theoretical papers have addressed the properties of
graphene with resonant impurities~\cite{ShytovPRL09, AbaninPRL10,
Cheianov09SolidS, Cheianov09PRB, CheianovEPL10, CheianovPRB11,
BaskoPRB08, WehlingChemPhysL09, WehlingPRB09, WehlingPRL10,
Titov10PRLa, Titov10PRLb, Titov11PRL, Mishchenko12, Titov13}.
Experimentally a way to create the strong atom-size small
impurities is by chemical functionalization of graphene by
impurity adatoms (see e.g.~Refs.~\cite{KimNanoLett08,
GeimScience09, BaldwinCarbon11, RobinsonNanoLettt10}). Mobility
inherent for adatoms allowed to put forward theoretical proposals,
suggesting the impurities sublattice-ordering caused by their
Casimir-like interaction~\cite{ShytovPRL09, AbaninPRL10,
Cheianov09SolidS, Cheianov09PRB, CheianovEPL10, CheianovPRB11}.
This ordering would lead to the controllable opening of the gap in
the electron's spectrum, highly desired for the graphene
electronics. Resonance at the Dirac point in this case is
necessary in order to make the Casimir interaction sufficiently
long-ranged.

The choice of the theoretical model describing the impurity
requires a special attention. Adding a large potential at a
certain carbon atom, a scheme assumed by most of the authors, does
not work~\cite{Mishchenko12}, since this requires the use of
unrealistically large impurity potential (hundreds ${\rm eV}$-s!).
A more realistic model, which is adopted in the current paper, was
suggested in Refs.~\cite{WehlingChemPhysL09, WehlingPRB09}, where
the adatom is treated not as a potential scatterer, but as a
quantum level tunnel-coupled to one of the carbon sites. The
tunnelling amplitude between the adatom and carbon atom is of
order $\sim 1{\rm eV}$, but the energy of the electron's level at
the impurity turns out to be very close to the Dirac point, thus
leading to the resonance scattering in undoped graphene. Density
functional theory analysis~\cite{WehlingChemPhysL09, WehlingPRB09,
WehlingPRL10} indicates that this situation may indeed be realized
for several kinds of impurity atoms.

In this paper we consider graphene functionalized by the impurity
atoms in a strong perpendicular magnetic field with the special
emphasis on the investigation of interaction between electrons
surrounding the impurities, a combination of problems never
touched in the existing literature. Several interesting and
potentially experimentally relevant results for both the electron
spectrum and the spin density caused by the joint effect of the
impurity and magnetic field will be presented.

A peculiar feature of graphene in a magnetic field is the
existence of zero-energy Landau levels with fully isospin
(sublattice) polarized electron states~\cite{SharapovPRL05}. Thus
in the presence of impurity one needs to analyze the coexistence
of two kinds of zero-energy states, localized due to the impurity
and due to the quantum cyclotron motion.

We use the model of Ref.~\cite{WehlingChemPhysL09} to describe the
impurity atoms and assume that the energy of the impurity level is
small compared to the Landau levels splitting, which means a
particle-hole symmetric limit. Both single- and many-impurities
problems, with and without electron-electron interaction will be
considered.

The paper starts with the investigation of the zero energy state
induced by a single impurity in the magnetic field in case of
non-interacting electrons. Remarkably in this case we found a
simple analytical solution for the wave function and energy. The
solution is approximate, with the small parameter being the
inverted large logarithm (of the Larmor radius divided by the
graphene lattice spacing), but it is sufficient to describe the
wave functions and the occupation of the eigenstates. The
important properties of the result are: First, the new resonant
states are indeed impurity-induced states in the graphene layer
and not the states localized on the impurity. The probability to
find the electron in this state in graphene plane increases in the
vicinity of the impurity, but the probability to stay exactly on
the impurity atom is parametrically small. Second, both the $n=0$
Landau level states and the impurity states without magnetic field
have exactly zero energies. That was the reason to expect both
kinds of states in undoped graphene to be half occupied and
spin-polarized~\cite{YazevPRL08, LehtinenPRL03, WehlingNano08}.
Now, when both the impurity and the magnetic field are present,
the impurity induced states acquire a finite energy and their
number is doubled, since the non-zero energy states may only
appear in pairs in the particle-hole symmetric limit. Doubling the
number of levels becomes possible because the impurity level in
the magnetic field gets hybridized with the $n=0$ Landau level
electron state most coupled to the impurity. The negative energy
impurity induced level is doubly occupied, thus reducing the total
polarization of the Quantum Hall ferromagnetic state of graphene
in a magnetic field.

A realistic description of the spin-polarization effects in
quantum Hall regime is impossible without taking into account
interaction between electrons. That is why, after solving the
noninteracting problem we proceed with the calculation of the
spectrum, taking into account the exchange electronic interaction
in the Hartree-Fock approximation. As expected, adding the
exchange interaction increases the Zeeman splitting of the $n=0$
Landau level in graphene by $1-2$ orders of magnitude. What is new
and interesting, we found that for the states surrounding the
impurity the splitting of up- and down-spin states depends on the
angular momentum number,~$m$. The electron with $m=1$, which is
not directly connected to the impurity gets the smallest
interaction induced Zeeman splitting and thus the smallest
excitation energy.

In this paper we treat the electron-electron interaction in the
Hartree-Fock approximation in the Hilbert space restricted to the
$n=0$ Landau level. Strictly speaking, this approach accounts
fully for the electron's interaction only in the first order. This
is justified because the electron-electron interaction in graphene
effectively is not very strong (and even may be further suppressed
by covering the sample by the material with large dielectric
constant~\cite{dielectric1,dielectric2,dielectric3,dielectric4}).
Moreover, all the interaction related predictions of this paper
are of qualitative nature and should remain intact after taking
into account higher order corrections in case of moderately strong
interaction.


In addition to the single impurity we consider the group of
several (many) impurity atoms attached to the graphene sheet in a
strong magnetic field. Existing publications investigate multiple
impurities distributed with a uniform density over the graphene
sample, leaving aside the potentially experimentally relevant case
of bunched impurities. To fill this gap we consider several
adatoms forming a dense bunch, such that the distance between any
two adatoms is small compared to the Larmor radius (but still
large compared to the lattice spacing). The Casimir interaction
between impurities~\cite{ShytovPRL09} favors the configurations
with the impurities coupled to the same sublattice of graphene
(Kekule ordering). Consequently we investigate the bunch of
impurities coupled all to the same sublattice of graphene. The
main result in this case is that a dense bunch of many
impurities/adatoms changes the electronic structure within a large
area of a graphene flake around it. As was written above, an
impurity in the magnetic field creates a couple of localized
states by hybridizing the singular $\sim 1/(x+iy)$ state of
Ref.~\cite{Pereira06} with the $n=0$ Landau level state mostly
connected to the impurity. In the case of many impurities, each of
them tends to create such a couple of hybridized levels, for which
it needs the $n=0$ Landau level state. Thus many $n=0$ states with
the angular momentum $m\ge 0$ become hybridized, leading to the
formation of a spin-unpolarized circular droplet of electrons
residing on one of the graphene sublattices around the bunch of
impurities. The energies of electrons, forming the droplet,
although nonzero, unlike the energies of other $n=0$ electrons,
decrease fast with the increasing angular momentum. Namely,
increasing $m$ by one leads to a decrease of the energy by a small
factor $\sim \langle r_{ab}\rangle/l$, where $\langle
r_{ab}\rangle$ is the typical distance between impurity atoms in
the bunch and $l$ is the Larmor radius.

Taking into account the exchange electron-electron interaction
changes the properties of the unpolarized electron droplet
surrounding the bunch of impurities in two important ways. First,
the energies of electrons forming the droplet now become all of
the same order of magnitude, $\sim e^2/l$. (The energies of
electrons with larger angular momentum $m$ are still smaller. But
the smallness is due to a pure numerical factor in the energy
$\eps_m$.) The second feature is that after the exchange
interaction is taken into account, electrons surrounding the
droplet start to feel the existence of electrons inside the
droplet. More precisely, as we will show, the electrons from the
$n=0$ Landau level staying outside the droplet interact via the
exchange interaction with each other, but not with the electrons
from the droplet. Thus the circumference of the droplet serves as
an edge for the outside electrons, leading to a circular edge
current.

Resonant impurities in graphene in a magnetic field were
considered recently in Ref.~\cite{Titov13}. However, this paper is
concentrated on transport properties and does not investigate the
electronic spectrum and spin structure. What is more important,
authors of Ref.~\cite{Titov13} do not consider electrons
interaction in graphene.

The paper is organized as follows. In Sec.~\ref{Section_Single} we
introduce the Hamiltonian for graphene and the attached impurity
atom and solve the single-impurity problem in the magnetic field
for non-interacting electrons. To do this we first present the
solution for Landau level states in graphene in the polar gauge.
The important supplementary information for
Sec.~\ref{Section_Single} is included in
Appendix~\ref{Appendix_A}. In Sec.~\ref{Section_Interaction} we
consider the single impurity problem with the electron-electron
interaction taken into account. The main part of the section is
devoted to the explanation of the structure of exchange
interaction in the case of the particle-hole symmetric Dirac
equation in graphene and to the discussion of how differently the
states with zero and non-zero energy experience the exchange
interaction in this case. Details of calculations for this section
are given explicitly in Appendix~\ref{Appendix_B}. In
Sec.~\ref{Section_Bunch} we consider the problem of many close
(closer than the Larmor radius) impurity atoms. The
subsection~\ref{Subsection_Bunch} deals with the exchange Coulomb
interaction in case of a bunch of impurities. Calculational
details for Sec.~\ref{Section_Bunch} are given in the
Appendices~\ref{Appendix_C} and~\ref{Appendix_D}.
Section~\ref{Section_Conclusions} presents the conclusions.

\section{Single impurity}\label{Section_Single}

A graphene plane with an impurity atom chemically bonded to it is
described by the Hamiltonian
 \bq\label{HamLattice}
H= t\sum_{\langle i,j\rangle} (a^{\dagger}_{i} b^{}_{j} +
b^{\dagger}_{j} a^{}_{i}) +U(a^{\dagger}_{0} d+d^\dagger a^{}_{0})
+\eps_d d^\dagger d,
 \ee
where operators $a_j^\dagger$, $b_j^\dagger$ create an electron on
the $j$-th site of the triangular sublattices $A$ and $B$ of the
honeycomb lattice of graphene, and $d^\dagger$ creates an electron
at the impurity atom attached to the carbon atom of the sublattice
$A$ with $j=0$. Electrons hop from a site of one sublattice to the
nearest sites of another sublattice with the matrix element
$t\approx 2.7eV$. Hopping matrix element between the impurity and
the carbon atom nearest to it has the same order of magnitude,
$U\sim t$. Crucial for us is the prediction of
Refs.~\cite{WehlingChemPhysL09, WehlingPRB09} that the energy of
the electron at the impurity may be anomalously small, $\eps_d\ll
t,U$, for several popular choices of the impurity atom. Only in
the case of the impurity level being very close to the Dirac
point, the model Eq.~(\ref{HamLattice}) leads to resonant
scattering of low energy electrons, which is a necessary
ingredient for Refs.~\cite{ShytovPRL09, AbaninPRL10,
Cheianov09SolidS, Cheianov09PRB, CheianovEPL10, CheianovPRB11,
BaskoPRB08, WehlingChemPhysL09, WehlingPRB09, WehlingPRL10,
Titov10PRLa, Titov10PRLb, Titov11PRL, Mishchenko12, Titov13}. That
is why in this paper we will always assume a negligibly small
impurity level energy, $\eps_d=0$.

Single-particle eigenstates of the Hamiltonian
Eq.~(\ref{HamLattice}) are created by the operator
 \bq\label{Phioperator}
\hat\Phi^\dagger =f_d d^\dagger + \sum_n f_n \hat\Psi_n^\dagger \
, \ \hat\Psi_n^\dagger =\sum_j (u_{nj}a^{\dagger}_{j}
 +v_{nj}b^{\dagger}_{j}).
 \ee
Here, $f_d$ is a probability amplitude to find the electron on the
impurity atom and $f_n$ are the probability amplitudes to find the
electron in the $n$-th eigenstate of the pure graphene
Hamiltonian. These eigenstates are determined by two complex
amplitudes $u_{nj}$ and $v_{nj}$ describing the electron residing
on one of the graphene sublattices, $A$ or $B$, in the unit cell
$j$. As usual, the same unit cell amplitudes $u_{nj}$ and $v_{nj}$
are combined into a (pseudo)spinor $\Psi_n(\vec r_j)$. The low
energy behavior of the wave-function $\Psi_n$ is captured by two
spinor envelope functions $\psi_n$ and~$\psi'_n$
 \bq\label{PsiUpper}
\Psi_n(\vec r_j) =\left(\begin{array}{cc}
u_{nj} \\
v_{nj}
\end{array}
\right) =e^{i{\bf Kr}_j}\psi_n(\vec r_j) + e^{i{\bf
K'r}_j}\psi_n'(\vec r_j),
 \ee
where vectors ${\bf K}$ and ${\bf K'}=-{\bf K}$ are directed to
two inequivalent corners of the Brillouin zone.

In order to proceed with solving the impurity problem in
Eqs.~(\ref{HamLattice},\ref{Phioperator}) one first needs to find
the eigenfunctions of the clean graphene Hamiltonian in a magnetic
field. The latter may be added into the Hamiltonian
Eq.~(\ref{HamLattice}) by introducing the coordinate dependent
hopping matrix elements $t_{ij}$ with properly chosen phases. We
however will be interested only in the low energy limit, when the
two envelope functions $\psi, \psi'$ become the eigenfunctions of
two decoupled massless Dirac Hamiltonians~\cite{review09} with the
magnetic field entering via the covariant derivative $\vec
p\rightarrow \vec p - (e/c)\vec A$. In the polar gauge, ${\bf
A}=({By}/{2}, \ -{Bx}/{2}, \ 0)$ the two Dirac Hamiltonians for
$K$ and $K'$ valleys become
 \bq\label{DiracB}
{\cal H}= i\eps_B
(\tau_- Q -\tau_+ Q^+) \ , \ {\cal H}'=i\eps_B
(\tau_- Q^+ -\tau_+ Q).
 \ee
Here $\tau_\pm =(\tau_x \pm i\tau_y)/2$ and $\tau_{x,y}$ are the
Pauli matrices in the pseudospin space, and $\eps_B=\hbar v_F/l$,
where $l=\sqrt{\hbar c/eB}$ is the Larmor radius. The Fermi
velocity in graphene is determined by the hopping matrix element
$t$ in Eq.~(\ref{HamLattice}) and the carbon-carbon distance $d$,
$\hbar v_F=3dt/2$. The creation and annihilation operators $Q^+$
and $Q$, in terms of the dimensionless complex coordinate
$z=(x+iy)/l$ have the form
 \bq
Q^+=\fr{1}{\sqrt{2}}\left( -2\fr{\partial}{\partial z^*}
+\fr{z}{2}\right) , \ Q=\fr{1}{\sqrt{2}}\left(
2\fr{\partial}{\partial z} +\fr{z^*}{2}\right)\ .
 \ee
They satisfy the ``oscillator" commutation relations $[Q,Q^+]=1$.
To find the eigenfunctions of ${\cal H}$ and ${\cal H}'$ we
introduce a set of (normalized) functions, $n,m>0$,
 \bq
\phi_{n,m-n}= \fr {(-{Q^+})^n{z^*}^m e^{-z z^*/4}}{\sqrt{2\pi n!
2^m m!}}\ .
 \ee
Here, the second index is responsible for the angular behavior,
$\phi_{n,k}\sim (z^*/|z|)^k$, and may be both positive and
negative. The two sets of eigenfunctions are now for $n>0$ (the
$n=0$ case should be considered separately)
 \begin{align}\label{psinm}
\psi_{\pm n,k} =
\left(\begin{array}{cc}
\fr{1}{\sqrt{2}}\phi_{n,k} \\
\fr{\mp i}{\sqrt{2}}\phi_{n-1,k+1}
\end{array}
\right),  \psi_{\pm n,k}'
=
\left(\begin{array}{cc}
\fr{1}{\sqrt{2}}\phi_{n-1,k+1} \\
\fr{\mp i}{\sqrt{2}}\phi_{n,k}
\end{array}
\right).
 \end{align}
The corresponding energies $\eps_{\pm n}$ and $\eps_{\pm n}'$
coincide and depend only on the Landau level number $n$
 \bq\label{energySingle}
\eps_{\pm n}=\eps_{\pm n}'=
\pm\sqrt{n}\eps_B.
 \ee
For each value of $n$ only two of the solutions Eq.~(\ref{psinm}),
$\psi_{\pm n,0}(z=0)= \psi_{\pm n,-1}'(z=0)= (1/\sqrt{2\pi},0)$,
have nonzero upper component at the origin and can be coupled to
the impurity attached to the carbon atom $A$ with $\vec r_j=0$.

The most interesting for us will be the zero energy states from
the $n=0$ Landau level. In undoped graphene one spin component of
this level is fully occupied by electrons and the other spin
component remains empty, leading to a strongly spin-polarized
state often called a Quantum Hall ferromagnet. Wave functions for
the $n=0$ Landau level have also a very special form, with
electrons from the $\vec K$ valley residing solely on the
sublattice $A$ and electrons from the $\vec K'$ valley on the
sublattice $B$. Corresponding envelope functions are
 \bq\label{eigenfunction0}
\psi_{0,m} =\left(\begin{array}{cc}
\phi_{0,m} \\
0
\end{array}
\right), \  \psi_{0,m}' =\left(\begin{array}{cc}
0 \\
\phi_{0,m}
\end{array}
\right).
 \ee
Only one state from the $n=0$ level, $\psi_{0,0}$, has a
nonvanishing upper component at $\vec r=0$ and can be coupled to
the impurity. All the other zeroth Landau level states are
decoupled from the carbon site connected to the impurity and thus
remains the exact zero energy eigenstates of the full Hamiltonian
Eq.~(\ref{HamLattice}).

Similarly, for each of the other Landau levels, $n\neq 0$, one may
choose a basis with only one state having non-vanishing
probability amplitude at the impurity. This single coupled state
has to be a superposition of solutions belonging to different
valleys, Eq.~(\ref{DiracB}), each being non-zero at the carbon
atom coupled to the impurity. Explicitly the subset of eigenmodes
of the graphene hopping Hamiltonian in the magnetic field coupled
to the impurity is, for $n\neq 0$,
 \begin{eqnarray}\label{eigenfunctions}
\Psi_{ n} =\fr{e^{i{\bf Kr}}}{2}\left(\begin{array}{cc}
\phi_{|n|,0} \\
\fr{n}{i|n|}\phi_{|n|-1,1}
\end{array}
\right)+\fr{e^{i{\bf K'r}}}{2}\left(\begin{array}{cc}
\phi_{|n|-1,0} \\
\fr{n}{i|n|}\phi_{|n|,-1}
\end{array}
\right).
 \end{eqnarray}
This set is completed by adding the $n=0$ solution
$\Psi_0=e^{i\vec K\vec r}\psi_{0,0}$ (\ref{eigenfunction0}).
(Another combination of two spinor functions
from~Eq.~(\ref{eigenfunctions}) with opposite relative sign
vanishes at the carbon site coupled to the impurity atom.)

All wave functions $\Psi_n$ Eq.~(\ref{eigenfunctions}) are equally
coupled to the impurity, $\Psi_n(0)= (1/\sqrt{2\pi},0)$. Thus the
problem of describing the effect of resonant impurity in graphene
reduces to solving the problem with a single impurity-level
equally coupled to a discrete ladder of special graphene states in
magnetic field. The sum over $n$ in Eq.~(\ref{Phioperator}) now
includes only the ladder states $\Psi_n$. The energy and the
complex amplitudes $f_n, f_d$ satisfy the equations
 \begin{eqnarray}\label{KatsMainMagnetic0}
(\eps-\eps_n)f_n=V_d f_d \ , \
(\eps-\eps_d)f_d=V_d\sum_n f_n \ .
 \end{eqnarray}
Here $V_d =3^{\fr{3}{4}} Ud/(2\pi^{\fr{1}{2}}l)$. When deriving
$V_d$ one should remember that the Hamiltonian
Eq.~(\ref{HamLattice}) acts in the lattice space, while the ladder
states Eq.~(\ref{eigenfunctions}) are the normalized functions of
the continuous coordinates. Eqs.~(\ref{KatsMainMagnetic0}) may be
rewritten as an algebraic equation for the energy and expression
for all amplitudes $f_n$ through a single ``impurity" amplitude
$f_d$,
 \begin{eqnarray}\label{KatsMainMagnetic}
\eps -\eps_d=\sum_n \fr{V_d^2}{\eps-\eps_n}\ , \
f_n=\fr{V_d}{\eps-\eps_n}f_d \ .
 \end{eqnarray}
The probability to find the electron on the impurity is found from
the normalization condition, yielding
 \begin{align}\label{KatsMainMagneticfd}
f_d^2= \fr{1}{1+V_d^2\sum 1/{(\eps-\eps_n)^{2}}}.
 \end{align}

\begin{figure}
\includegraphics[width=9.cm]{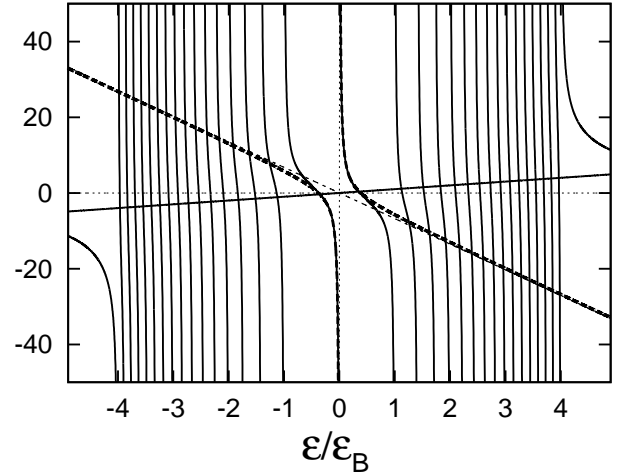}
\caption{Finding graphically the energy from the first equation
(\ref{KatsMainMagnetic}). (Energy in units of $\eps_B$ and
$V_d=1$.) Straight solid line with small positive slope show the
left hand side of the equation. Other solid lines show the right
hand side of Eq.~(\ref{KatsMainMagnetic}), having a pole at
$\eps=0$ and other poles at each $\eps=\pm\sqrt{n}\eps_B$. Each
crossing of two solid lines corresponds to an energy level.
Impurity induced localized states $\Psi_{S_\pm}$ correspond to two
crossings closest to the zero energy. In addition there are two
crossings far to the right and far to the left from the energy
segment shown in the figure (well outside the energy band of
graphene). These crossings correspond to the bonding and
anti-bonding states of the impurity atom and a single carbon atom
closest to it. They are responsible for the creation of the true
chemical bond between the two atoms. Thick dashed line with a
single pole at $\eps =0$ shows the approximate form of the energy
equation~(\ref{KatsMainMagnetic}) r.h.s. described in the
Appendix~\ref{Appendix_A}.} \label{figEnergy}
\end{figure}

A graphical solution of the energy equation
Eq.~(\ref{KatsMainMagnetic}) is shown in figure~\ref{figEnergy}.
Energy levels are the energies at which the straight solid line
representing the left hand side of the equation crosses the
multiple solid lines showing the sum of hyperbolae
$1/(\eps-\eps_n)$ in the right hand side. The impurity-induced
states correspond to the two crossings most close to zero energy.

\begin{figure}
\includegraphics[width=9.cm]{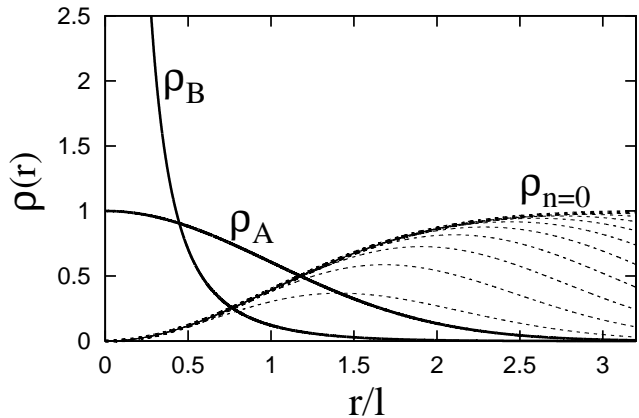}
\caption{Radial dependance of the electron density at two
sublattices, $\rho_A$ and $\rho_B$, for the impurity induced state
$\Psi_{S_-}$ (\ref{singularB}). The density is doubled due to two
spin components and normalized such that a single occupied Landau
level has $\rho=1$. For drawing we choose $L=\ln (l/d)=5$. The
thick dashed line shows the density $\rho_{n=0}$ of the $n=0$
Landau level from sublattice $A$, valley $K$, with one spin
orientation and having angular momenta $m=1,2,3,\cdots$ (the
states not affected by the impurity). Thin dashed lines show, how
this density is build by adding one by one electrons with $m=1$,
$m=2$, etc.. Together $\rho_A$ and $\rho_{n=0}$ give a constant
charge density of a fully occupied Landau level. However, these
constant density electrons have a very nontrivial exchange
interaction, discussed in Sec.~\ref{Section_Interaction}. The
singular increase of density in state $\Psi_S$ at sublattice $B$,
$\rho_B\sim 1/r^2$, is fully compensated by the decrease of
density in Landau levels with $n>0$, as follows from the
particle-hole symmetry. } \label{figDensity}
\end{figure}

Details of the explicit analytical solution of
Eqs.~(\ref{KatsMainMagnetic}) are given in the
Appendix~\ref{Appendix_A}. Here we show only the results for the
in-plane wave function, $\Psi_S=\sum f_n\Psi_n$, and energy of the
singular impurity-induced state.
 \bq\label{singularB}
\Psi_{S_\pm} =
\left(\begin{array}{cc}
 e^{i{\bf Kr}}{\displaystyle \fr{e^{-z z^*/4}}{2\sqrt{\pi}}} \\
 \pm i
 {\rm Im}{\displaystyle[ \fr{e^{i{\bf Kr}}}{z}
]}{\displaystyle \fr{e^{-z z^*/4}}{\sqrt{2\pi L}}}
\end{array}
\right), \ \eps_{S_\pm}=\fr{\pm\eps_B}{2\sqrt{L}} \ .
 \ee
The analytical solution is found in the large logarithm limit,
$L=\ln (l/d)\gg 1$, {\it i.e.} in case of Larmor radius $l$ very
much exceeding the carbon-carbon distance $d$. However, the
resulting approximate wave function allows to extract many
qualitative features of the exact solution. As it should be for
the eigenfunctions of the particle-hole symmetric Hamiltonian
Eq.~(\ref{HamLattice}) for vanishing $\eps_d$, there are two
solutions, $\Psi_{S_\pm}$, with opposite signs of energy. Electron
in the state described by Eq.~(\ref{singularB}) can hop to the
impurity attached at $\vec r=0$ to the carbon atom from the
sublattice $A$. However, the probability to find an electron on
the impurity, found via Eq.~(\ref{KatsMainMagneticfd}) is small,
$f_d^2=\sqrt{3}\pi t^2/(4U^2 L)\ll 1$, and most of the time the
electron spends in the graphene plane (remember that $U\sim t$).
The electronic density corresponding to Eq.~(\ref{singularB}) is
shown in Fig.~\ref{figDensity}.

The wave functions $\Psi_{S_\pm}$ Eq.~(\ref{singularB}) have a
clear physical interpretation. The two states $\Psi_{S_\pm}$ may
be thought of as equal weight superpositions of two simple states,
which contribute to their upper and lower components,
respectively. First is the $n=0$ Landau state $\psi_{0,0}$
Eq.~(\ref{eigenfunction0}), having only one nonvanishing component
(upper). The lower component of Eq.~(\ref{singularB}) comes from
the zero energy localized state found in Ref.~\cite{Pereira06}
with the large distance cutoff at the Larmor radius, $r\sim l$.
Note that the lower component of $\Psi_{S_\pm}$ contain similar
contributions from both the $K$ and $K'$ valleys.

The energies of the states $\Psi_{S_\pm}$ in Eq.~(\ref{singularB})
are also small compared to the Landau level splitting $\eps_B$,
but only as an inverse square root of the large logarithm. A
chemical potential in undoped graphene coincides with the Dirac
point. This suggests that both spin components of the level
$\eps_{S_-}$ are occupied and form a spin singlet, thus preserving
the particle-hole symmetry and electrical neutrality of the
system, graphene plus impurity. Taking into account spin of the
electron leads also to a small Zeeman splitting of all energy
levels introduced in this section. Without the impurity, the
simplest choice of occupation of Landau levels in graphene is to
fully occupy both $K$ and $K'$ valley components of the $n=0$
level with spin down while keeping empty their spin up
counterpart. This Quantum Hall ferromagnetic state minimizes the
Zeeman energy of half occupied $n=0$ Landau level in neutral
undoped graphene. Thus according to Eq.~(\ref{singularB}) each
adatom tends to lower the polarization of the Quantum Hall
ferromagnet by one electron spin i.e. by $1/2$. As will be shown
in the following section, this scheme of occupying the $n=0$
Landau level remains intact after taking into account
electron-electron interactions.

\section{Single impurity with
interaction}\label{Section_Interaction}

A bare Zeeman splitting for electrons in graphene is linear in
magnetic field, $E_Z=g \mu_B B$. The splitting of Landau levels,
$\eps_B\propto \sqrt{B}$, scales as the square root of the
magnetic field and formally grows much slower with the increasing
field than $E_z$. However, for any realistic values of the
magnetic field, the Zeeman splitting is negligibly small compared
to the inter-level distance. E.g., for the magnetic field $10$T,
$\eps_B/E_Z \approx 70$~\cite{AbaninPRL06}, while the typical
interaction energy $e^2/l \sim (e^2/\hbar v_F)\eps_B\sim \eps_B$.
Thus similarly to what happens in the conventional semiconductor
heterostructures, the spin physics in magnetic field in graphene
is dominated by the electronic interaction. The strong
renormalization of the Zeeman splitting is seen already in the
Hartree-Fock approximation and is due to the exchange interaction.
Direct interaction, i.e. Hartree, is trivial even in the presence
of the impurity atom, since we are working in the particle-hole
symmetric limit (see Fig.~\ref{figDensity} and discussion in the
caption).

\begin{figure}
\vspace{1.cm}
\includegraphics[width=8.cm]{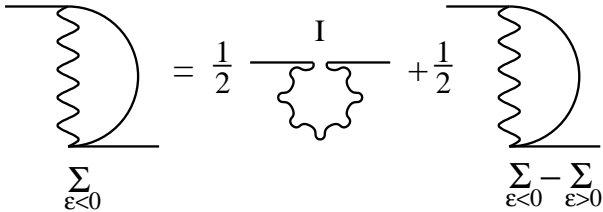}
\caption{Diagrammatic presentation of the exchange energy. Wavy
line shows the Coulomb interaction $e^2/|{\vec  r}-{\vec r}'|$.
Summation over the complete set of states in the first term in
r.h.s. gives unity, or $\delta$-function. }\label{fig.Exchange}
\end{figure}

We begin this section with showing that due to the particle-hole
symmetry of the Dirac equation, the exchange interaction acts very
differently on the electrons from the $n=0$ and $n\neq 0$ Landau
levels in graphene. To calculate the exchange energy of the
electron in a state~$\Psi$,
 \bq\label{exchangeGeneral}
\eps_{exch}=-\int \Psi^\dagger(\vec r)\fr{e^2}{|\vec r-\vec
r'|}\rho(\vec r,\vec r')\Psi(\vec r')d\vec r d\vec r',
 \ee
one needs to find the projection operator onto the occupied states
 \bq\label{rho}
\rho(\vec r,\vec r')= \sum_n \nu_n\Psi^{}_n(\vec
r)\Psi_n^\dagger(\vec r'),
 \ee
where $\nu_n$ is the occupation number for the state $n$, taking
values $0$ or $1$ (we consider only the zero temperature limit in
this paper). In case of undoped graphene with $\eps_F=0$ it is
convenient to split the summation in Eq.~(\ref{rho}) into a sum
over the complete set of all states and a difference of two sums
over occupied and empty states, leading to
 \begin{eqnarray}\label{funny}
\rho(\vec r,\vec r')=\fr{\bf I}{2} +\sum_i
\fr{2\nu_i-1}{2}\Psi_i(\vec r)\Psi_i^\dagger(\vec r').
 \end{eqnarray}
The unity operator ${\bf I}\sim \delta(r-r')$ here causes the
uniform shift of all on-site energies in the Hamiltonian
Eq.~(\ref{HamLattice}) (see Fig.~\ref{fig.Exchange}) and may be
ignored.

Since the signs of two sublattice components of the graphene
Hamiltonian eigenfunctions may always be chosen as
$u_{\eps}=u_{-\eps}$ and $v_{\eps}=-v_{-\eps}$, the non-zero
energy contribution to the r.h.s. of Eq.~(\ref{funny}) takes the
form
 \bq\label{exchangeMatrix}
\sum_{\eps_i\neq 0} \fr{2\nu_i-1}{2}\Psi_i\Psi_i^\dagger
=\sum_{\eps_i<0}
\left(\begin{array}{cc} \ \ 0 \ \ , u_{i}v_{i}^*  \\
\!\!\! v_{i}u_{i}^*  , \ \ 0
\end{array}\right).
 \ee
On the other hand, the zero energy eigenfunctions
Eq.~(\ref{eigenfunction0}) of the clean graphene Hamiltonian
reside solely on one sublattice and their contribution to the
projection operator can only be a diagonal matrix
 \bq\label{exchangeMatriD}
\sum_{\eps_i= 0} \fr{2\nu_i-1}{2}\Psi_i\Psi_i^\dagger
=\sum_{\eps_i=0}\fr{2\nu_i-1}{2}
\left(\begin{array}{cc} \!\!\!\!\! u_{i}u_{i}^*  , \ \ 0 \\
 \ \ 0 \  , v_{i}v_{i}^*
\end{array}\right).
 \ee
For each state $i$ here only one component $u_i$ or $v_i$ differs
from zero~(\ref{eigenfunction0}). Now one easily sees that
electrons with $\eps=0$ interact via exchange only with other
$\eps=0$ electrons. On the contrary, electrons with $\eps\neq 0$
interact via exchange with the states with any energy.

Eqs.~(\ref{eigenfunction0}),(\ref{exchangeMatriD}) allow to
reproduce the known result~\cite{Jancovici81,Gusynin02} for the
exchange dominated Zeeman splitting of the $n=0$ Landau level (see
Appendix {\bf B})
 \bq\label{exchangeBareL}
\eps_{n=0}= \pm \left( \fr {e^2}{2l}\sqrt{\fr{\pi}{2}}
+E_Z\right).
 \ee
This exchange renormalized Zeeman energy is of the order of the
Landau levels interval, $\eps_B$, since in graphene $e^2/(\hbar
v_F)\sim 1$. The two signs of $\eps_{n=0}$ correspond to two
spin-projections on the magnetic field axis. We keep the small
$E_Z$ in Eq.~(\ref{exchangeBareL}) to compare with
Eq.~(\ref{exchangeS}) below.

Due to their nonzero energy, the exchange interaction for the
impurity induced localized states $\Psi_{S_\pm}$,
Eq.~(\ref{singularB}), is calculated completely differently, even
though they contain a one half admixture of the $n=0$ Landau level
state $\psi_{0,0}$, Eq.~(\ref{eigenfunction0}). In the large
logarithm limit, $L=\ln(l/d)\gg 1$, after the subtraction of the
uniform energy shift due to the unity operator in
Eq.~(\ref{funny}), the impurity states $\Psi_{S_\pm}$ interact via
exchange only with themselves via the nondiagonal density operator
Eq.~(\ref{exchangeMatrix}). The energies of two lowest (occupied)
of these levels are
 \bq\label{exchangeS}
\eps_{S_-}=-
\fr{\eps_B}{2\sqrt{L}} -\fr{e^2}{2l}\sqrt{\fr{\pi}{2}}
\pm E_Z .
 \ee
The first term here is the noninteracting energy
Eq.~(\ref{singularB}), which is suppressed by the inverse square
root of the large logarithm. The largest second term is the
exchange energy, which occasionally turns out to be the same as
the exchange energy for the $n=0$ Landau levels without impurity,
Eq.~(\ref{exchangeBareL}). For undoped graphene both spin/Zeeman
components of the level Eq.~(\ref{exchangeS}) are occupied, thus
forming a singlet and reducing the total spin of the
quantum-Hall-ferromagnet state.

\begin{figure}
\includegraphics[width=6.5cm]{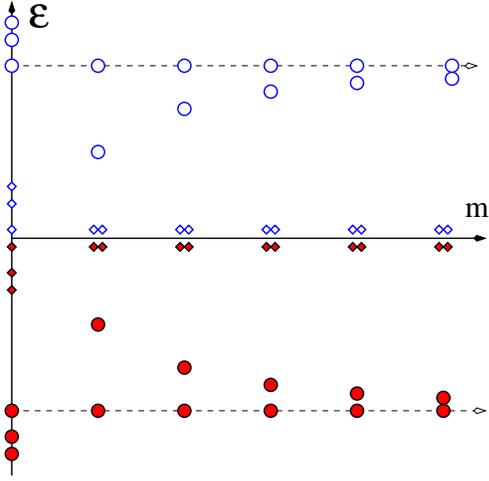}
\caption{Energies (schematic) of the $n=0$ Landau level states
around the impurity with and without electrons interaction. {\bf
Small rhombuses:} Show the energy levels for the noninteracting
problem, filled red rhombuses for filled and empty blue rhombuses
for empty states. For each $m>0$ there are two valley degenerate
states ($K$ and $K'$), each with two spin-components, split due to
the small Zeeman energy $\pm E_Z$. For $m=0$ there are total of
six low energy states. A pair of levels: spin split $n=0$, $m=0$,
$K'$-valley level behaves exactly the same way as its $m\neq 0$
companions. In addition there is a pair of spin-split low energy
levels $\Psi_{S_\pm}$, $\eps_{S_\pm}$, Eq.~(\ref{singularB}). Both
spin-components of $\eps_{S_-}$ are occupied, forming a singlet.
{\bf Big circles:} Show the energy levels for interacting
electrons. Again filled red circles show occupied states and empty
blue circles show unoccupied states. States from the $K'$ valley
are not affected by the impurity independent on the value of $m$.
Their energies are given by Eq.~(\ref{exchangeBareL}), which is
shown by two dashed lines. Occasionally, the exchange interaction
causes the same shift of the $\eps_{S_\pm}$ levels, as it did for
$n=0$ Landau level states in the absence of impurity. These states
are shown by the two lowest red circles and by the two highest
empty blue circles at $m=0$. The most interesting are the states
with small, but finite values of the angular momentum number $m$
from the $K$ valley. Due to their reduced exchange interaction
these states fall inside the gap between the usual spin up and
down $n=0$ Landau levels shown in Eq.~(\ref{exchangeBareL}). }
\end{figure}

The Zeeman splitting of the $n=0$ Landau level enhanced by the
exchange interaction, Eq.~(\ref{exchangeBareL}), does not depend
on the angular momentum quantum number $m$. This is very natural,
since our special choice of the vector potential, which made $m$ a
good quantum number, introduces only a spurious breaking of the
translational invariance in physically homogeneous systems. Adding
the impurity atom breaks the translational invariance and the
electron energy may now depend on angular momentum. This however
does not happen in the non-interacting case of the previous
section, where only the electrons with $m=0$ were affected by the
point impurity. The actual dependence on the angular momentum
appears only after taking into account the exchange interaction.

The mechanism leading to the $m$-dependance of the energy levels
is also interesting and relies on the difference between zero and
non-zero energy eigenstates of the particle-hole symmetric Dirac
equation discussed above. As we have shown, electrons from the
$n=0$ Landau level interact only with themselves via the exchange
interaction. For example, any $n=0$ electron from the $K'$ valley
($\psi'_{0,m}$ in Eq.~(\ref{eigenfunction0})) interacts with the
fully occupied $n=0$ level for the electron with spin down, or
with the completely empty $n=0$ level for the electron with spin
up. This leads to the energy Eq.~(\ref{exchangeBareL}). Note that
after the subtraction of the constant energy shift due to a
contact term in Eq.~(\ref{funny}), there is a nontrivial
interaction with both occupied and empty states.

On the other hand, the electron with spin down from the $K$-valley
and with $m\neq 0$ does not see the exchange attraction from the
same level with $m=0$, which was taken to build the finite energy
states $\Psi_{S_\pm}$ Eq.~(\ref{singularB}). Thus the missing
level pushes up the energies of levels around it. This effect is
strongest for the closest to the adatom electron with $m=1$, which
become the easiest electron to excite. The upward shift of the
levels become smaller with increasing angular momentum.
Similarly, for the empty subband of the $n=0$ Landau level from
the $K$-valley and with spin up, the absence of the level with
$m=0$ in the sum for the projection operator
Eq.~(\ref{exchangeMatriD}) lowers the energies of the states with
$m=1,2,\cdots$. Explicit calculation of the energy levels with
$m\ge 1$, leading to
 \bq\label{exchangeSm}
\eps_{0,m} =\pm \fr{e^2\sqrt{\pi}}{2\sqrt{2}l}\left(1-
\fr{\sqrt{2}(2m)!}{8^m(m!)^2} \right),
 \ee
is given in the Appendix~\ref{Appendix_B}. Energies of electron
states at and around the impurity are depicted on the figure 4.


\section{Bunch of impurities}\label{Section_Bunch}

In this section we consider a bunch of many closely spaced
impurity atoms coupled all to the carbon atoms from the same
sublattice $A$. The choice of fully sublattice-polarized bunch is
motivated by Refs.~\cite{ShytovPRL09, AbaninPRL10,
Cheianov09SolidS, Cheianov09PRB, CheianovEPL10, CheianovPRB11},
where it was shown that the Casimir interaction between
impurities, caused by electrons in graphene,  favors the
sublattice ordering.

By closely spaced impurities we mean close compared to the Larmor
radius $l$, but not as close as the carbon-carbon distance on the
hexagonal lattice, $d$. Instead of Eq.~(\ref{KatsMainMagnetic0})
we now write
 \begin{eqnarray}\label{KatsMainMagnetic2}
(\eps-\eps_n)f_n = \sum_a V_{na}f_a \ , \
(\eps-\eps_a)f_a = \sum_n V_{an}f_n.
 \end{eqnarray}
Here $f_a$ are the probability amplitudes to find the electron on
the impurity $a$ and $f_n$ is the amplitude to find the electron
in state $n$ in graphene plane (compare to
Eq.~(\ref{Phioperator})). The impurity onsite energies, $\eps_a$,
are assumed to be negligibly small.

In case of one impurity we were able to choose a single state
$\Psi_n$~(\ref{eigenfunctions}) from each Landau level, coupled to
the impurity by a uniform matrix element
$V_d$~(\ref{KatsMainMagnetic0}). For several impurities summation
over $n$ in Eq.~(\ref{KatsMainMagnetic2}) includes both summation
over the Landau levels and over the many individual states at each
Landau level. Matrix elements $V_{an}=V_{na}^*$ are now
proportional to the value of the upper component of the particular
electron's wave function at the carbon site coupled to the
individual impurity.

Let $N\gg 1$ be the number of the impurity atoms. First equation
(\ref{KatsMainMagnetic2}) expresses an infinite number of in-plane
states amplitudes $f_n$ through the $N$ impurity amplitudes~$f_a$
 \bq\label{fnN}
f_n = \fr{1}{\eps-\eps_n}\sum_a V_{na}f_a.
 \ee
The energy $\eps$ and impurity amplitudes $f_a$ should then be
found from the set of $N$ linear equations
 \begin{eqnarray}\label{GabN0}
(\eps-\eps_a)f_a = \sum_b G_{ab}f_b \ , \ G_{ab} =\sum_n
\fr{V_{an}V_{nb}}{\eps -\eps_n}.
 \end{eqnarray}
Equations (\ref{KatsMainMagnetic2}) and (\ref{GabN0}) are exact.
Explicit compact formula for the matrix $G_{ab}$ at low energies
$\eps$ is found in Eq.~(\ref{GabN2}) of Appendix~\ref{Appendix_C}.
The energies of impurity induced states are then estimated as
(each energy comes in a plus and minus pair)
 \bq\label{Hierarchy}
\eps_S^{(1)}=\fr{\eps_B}{2\sqrt{L}}, \ \eps_S^{(2)}\sim
\eps_B\fr{\langle |{\bf r}_{ab}|\rangle}{l}, \ \eps_S^{(3)}\sim
\eps_B\fr{\langle {\bf r}_{ab}^2\rangle}{l^2}, \ \cdots .
 \ee
Here $L=\ln ({\langle |{\bf r}_{ab}|\rangle}/{l})\gg 1$, $\langle
|{\bf r}_{ab}|\rangle $ is a typical distance between adatoms in a
bunch, $\langle {\bf r}_{ab}^2\rangle$ is a typical squared
distance between adatoms and so on. As we see, only one
eigenvalue, $\eps_S^{(1)}$, remains (almost) the same as it was in
case of a single impurity Eq.~(\ref{singularB}). Each next energy
is by a factor $\langle |{\bf r}_{ab}|\rangle/l\ll 1$ smaller than
the previous one. The reason for such hierarchy of energy
eigenvalues will be clear after considering the corresponding wave
functions.

Eqs.~(\ref{KatsMainMagnetic}) and (\ref{fnN}) show that the
electron wave function in graphene plane in case of multiple
impurities is a superposition of $N$ single impurity solutions
$\Psi_S$, Eq.~(\ref{singularB}). Since the impurities are very
close, the resulting sum of the wave functions may differ strongly
(and is different due to the severe cancellations) from any
individual contribution. Among $N$ wave functions described by
Eq.~(\ref{GabN0}) only one having the largest energy looks close
to the single impurity solution Eq.~(\ref{singularB}) (See
Appendix, Eq.~(\ref{singularBN0})). Wave functions of other
impurity induced states, having smaller energies $\eps^{(2)}$,
$\eps^{(3)}$, etc. in Eq.~(\ref{Hierarchy}), far away from the
small bunch of impurities have the form
 \bq\label{singularBN}
\Psi_S^{(m)} =
\left(\begin{array}{cc}
 e^{i{\bf Kr}}{\displaystyle\fr{\phi_{0,m}}{\sqrt{2}}}
 \\
 \pm e^{i\alpha}
 {\displaystyle\fr{e^{i{\bf K'r}}}{z^*}}
 {\displaystyle \fr{e^{-zz^*/4}}{2\sqrt{\pi L}}}
\end{array}
\right).
 \ee
Here $\alpha$ is some unknown unimportant constant. Details of
calculation of $\Psi_S^{(m)}$ are given in the
Appendix~\ref{Appendix_C}.

The accuracy of the wave function Eq.~(\ref{singularBN}) is much
better than the accuracy of the individual energies
Eq.~(\ref{Hierarchy}). The energies $\eps_S^{(m)}$ are known only
by the order of magnitude. The overall normalization of the
components of Eq.~(\ref{singularBN}) is found with $\sim 1/\log$
accuracy, but the accuracy of e.g. the upper component itself is
much better, being determined by the small ratio of the size of
the bunch and the Larmor radius.

\begin{figure}
\includegraphics[width=8.cm]{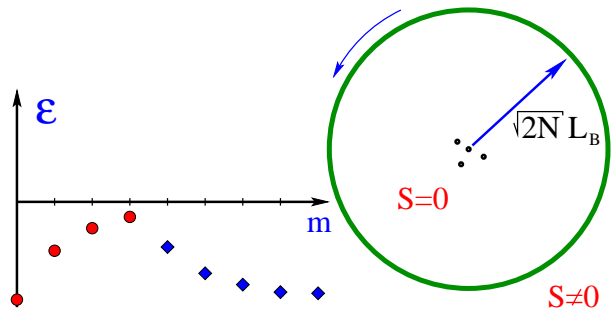}
\caption{Right: A large droplet of reduced spin polarization
surrounding a bunch of impurities. Left: Schematic angular
momentum dependance of the energy of impurity induced states (red
circles) and the energy of the $n=0$ sublattice $A$ electron
states surrounding the droplet (blue rhombuses). Due to their
nonuniform exchange energy, these latter states carry an edge
current around the droplet. }\label{fig.droplet}
\end{figure}

The lower components of the wave functions $\Psi_S^{(m)}$
Eq.~(\ref{singularBN}) are similar for any $m$. As was already
mentioned, Eq.~(\ref{singularBN}) appears as a result of strong
cancellations between the single-impurity solutions generated by
very close impurities. This cancellation suppresses the upper
component of $\Psi_S^{(m)}$, but also a part of the lower
component oscillating in unison with the upper one, $\propto
e^{i\vec K \vec r}$. The part of the lower component behaving like
$e^{i\vec K' \vec r}/z^*$ survives and is shown in
Eq.~(\ref{singularBN}).

More interesting is the upper component of the impurity induced
state $\Psi_S^{(m)}$. As was discussed in
Section~\ref{Section_Single}, effectively the impurity state
$\Psi_S$~(\ref{singularB}) is a superposition of the $n=0$ Landau
level state $\psi_{0,0}$~(\ref{eigenfunction0}) and the $\sim 1/z$
localized state of Ref.~\cite{Pereira06}. Each of them has only
one non-vanishing component, up or down. Consequently, the upper
component of the $N$-impurities solution $\Psi_S^{(m)}$ represents
a superposition of $N$ such zeroth Landau level solutions, each
centered at individual impurity. However, these spatially slightly
offset eigenfunctions of the graphene Hamiltonian in the magnetic
field are very similar. As is shown in the
Appendix~\ref{Appendix_C}, the upper components of the true
solutions of the many impurities problem Eq.~(\ref{GabN0}) became
the orthogonalized combinations of the $n=0$ Landau level states
generated by the individual impurities and these new orthogonal
states have an (almost) well defined value of the angular momentum
quantum number $m$. It is not surprising that the states
$\Psi_S^{(m)}$~(\ref{singularBN}) with large $m$ have small
energies Eq.~(\ref{Hierarchy}), since their coupled to the
impurity atom upper component vanishes at small distances as
$\phi_{0,m}\sim z^{*m}$.

Electrons
described by Eq.~(\ref{singularBN}) have a probability $1/2$ to be
found in the $n=0$ Landau level state in the $K$ valley. To
preserve the electro-neutrality of undoped graphene each such
$n=0$ state should be on average occupied by a one electron, which
is achieved if both spin components of the negative energy states
$\Psi_S^{(m)}$~(\ref{singularBN}) are occupied. Thus we expect the
electrons from the $K$ valley $n=0$ Landau level surrounding bunch
of impurities to form a large (radius $R=\sqrt{2N}l$) unpolarized
droplet inside the spin-polarized Quantum Hall ferromagnet, as is
shown in Fig.~\ref{fig.droplet}. The $n=0$ Landau level electrons
from the $K'$ valley remains fully spin-polarized.

\subsection{Many impurities with interaction}\label{Subsection_Bunch}

As was shown in section~\ref{Section_Interaction}, energies of
electrons surrounding the single impurity atom are strongly
modified by the electron-electron interaction. The same is true in
case of multiple impurities. The simple calculation sketched in
appendix~\ref{Appendix_D} gives the energies of impurity-induced
states Eq.~(\ref{singularBN})
 \bq\label{exchangeSFN}
\eps_S^{(m)}=\pm\fr{(2m)!}{(m!)^2
2^{2m}}\sqrt{\fr{\pi}{2}}\fr{e^2}{2l} \ .
 \ee
Unlike the noninteracting result Eq.~(\ref{Hierarchy}), all
energies Eq.~(\ref{exchangeSFN}) are of the same formal order of
magnitude, $\sim e^2/l$. The decrease of $\eps_S^{(m)}$ with
increasing $m$ is now due to a pure numerical factor. Although the
electronic interaction renormalizes strongly the energy
eigenvalues, wave functions Eq.~(\ref{singularBN}) are intact and
determine the dynamics of electrons surrounding the bunch of
adatoms.

The particle hole symmetry ensures that the occupied electron
states $\Psi_S^{(m)}$~(\ref{singularBN}) with $0\le m <N$ and the
Landau level states $\psi_{0,m}$~(\ref{eigenfunction0}) with $m\ge
N$ create a constant charge density around the bunch of adatoms.
This means that the direct electrons interaction is trivial.
However, as follows from
Eqs.~(\ref{exchangeGeneral},\ref{exchangeMatrix},\ref{exchangeMatriD}),
the states $\Psi_S^{(m)}$ and $\psi_{0,m}$ do not interact via the
exchange interaction. Consequently, the electrons with $m>N$ see
the weaker exchange interaction, since they are missing the
interaction with the electrons from inside the circle,
$r<l\sqrt{2N}$. Energies of these states, occupied or empty, may
be written as
 \bq\label{energyRing}
\eps_{0,m>N}=\mp\fr{e^2}{2l}\sqrt{\fr{\pi}{2}} \, g(m-N)\ ,
 \ee
where the function $g(x)$ increases smoothly from $g(0)=1/2$ to
$g(x\gg\sqrt{N})=1$. Explicit form of $g(x)$ could not be found in
a compact form, but some steps towards its evaluation are given in
appendix~\ref{Appendix_D}.

The electron density for each state $\psi_{0,m}$ has a form of the
narrow ring with the radius $R_m=l\sqrt{2m}$. The $m$-dependance
of the energies $\eps_{0,m>N}$~(\ref{energyRing}) of these ring
states means that there is an edge state carrying a persistent
current around the spinless electron droplet.


\section{Conclusions}\label{Section_Conclusions}

The aim of this paper was to investigate the effect of resonant
adatom impurities on the spectrum and electronic wave functions in
graphene in quantizing magnetic field. Despite the broad interest
in properties of graphene with resonant
impurities~\cite{ShytovPRL09, AbaninPRL10, Cheianov09SolidS,
Cheianov09PRB, CheianovEPL10, CheianovPRB11, BaskoPRB08,
WehlingChemPhysL09, WehlingPRB09, WehlingPRL10, Titov10PRLa,
Titov10PRLb, Titov11PRL, Mishchenko12, Titov13}, almost nobody up
to now (except for Ref.~\cite{Titov13} in our list) have
considered their role in the Quantum Hall regime. This gap needed
to be closed simply because of the experiments in magnetic field
are much easier in graphene than in other two-dimensional
materials~\cite{Novosel05,Zhang05}. Luckily we also found several
interesting effects, especially related to the electrons
interaction, which may potentially be observed in functionalized
graphene in this regime.

The resonance condition requires the impurity atom having an
electron level with the energy very close to the Dirac point in
graphene and the belief in the existence of such impurities is
based on the theoretical calculations of
Refs.~\cite{WehlingChemPhysL09, WehlingPRB09, WehlingPRL10}.
Experimental confirmation of these predictions, or even a direct
search for the materials having the proper resonant levels is thus
highly desired. Consequently, the direct measurement of the
spectrum of electrons surrounding the impurity,
Eq.~(\ref{exchangeSm}), may serve as a proof of the resonance. The
apparatus necessary for such experiments was already developed in
Ref.~\cite{EvaAndrei}, where the redistribution of the Landau
levels occupation around the charged impurity was measured with
the scanning tunnelling microscope.

One may also consider the same model with the impurity energy
$\eps_d$ shifted away from the neutrality point. The symmetry
between the positive and negative energy solutions of the Dirac
equation will then be broken, leading to the in-plane charge
redistribution and broken charge neutrality. The induced charges
however, will create the electrostatic potential pushing the
impurity level back to the Dirac point. Investigation of such
self-consistent stabilization of the model Eq.~(\ref{HamLattice})
with zero $\eps_d$ is an interesting problem for a future
investigation.

In this paper we were able to successfully describe not only
graphene with a single impurity, but also made a considerable
progress in solving the problem with many impurities forming a
small bunch. Such a strongly inhomogeneous distribution of atomic
impurities, which was never considered theoretically, is the
(likely) possible outcome of experimental functionalization of
graphene. According to our prediction, the bunch of impurities
creates a large droplet of reduced spin inside a Quantum Hall
ferromagnet, encircled by a current carrying edge state. Similarly
to the electron states induced by the single impurity, such
droplets should be measurable by the scanning tunnelling
microscope. In addition, due to its large size, even a single such
droplet may affect the transport in graphene mesoscopic devices.

\begin{acknowledgments}

Discussions with P.W.~Brouwer, E.~Kandelaki, P.~Recher,
M.~Schneider, G.~Schwiete and L.~Weithofer are greatly
appreciated. The author acknowledges hospitality of the Institute
for Advanced Studies at the Hebrew University, Jerusalem. This
work was supported by the Alexander von Humboldt Foundation and by
the DFG grant RE~2978/1-1.
\end{acknowledgments}


\appendix

\section{Solving the Dirac equation with a single
impurity}\label{Appendix_A}

In this Appendix we present the solution of
Eq.~(\ref{KatsMainMagnetic}) for the energy of the impurity
induced localized state in magnetic field. We will then find the
explicit shape of the localized state in the coordinate
representation Eq.~(\ref{singularB}).

Impurity induced localized states, which are the solutions of
Eq.~(\ref{KatsMainMagnetic}) most close to the Dirac point, have
$|\eps|\ll\eps_B$. Therefore it is convenient to consider
separately the $n=0$ and $n\neq 0$ contributions to the sum in the
right hand side of Eq.~(\ref{KatsMainMagnetic}). For $n\neq 0$ the
contributions with $n>0$ and $n<0$ almost cancel each other. It is
enough to consider the result of this cancellation in the linear
in $\eps$ approximation. With the energies $\eps_n$ found in
Eq.~(\ref{energySingle}) one has
 \bq\label{KatsMainApproximate}
\eps -\eps_d=\fr{V_d^2}{\eps}- \sum_{n=1}^{n_{\rm max}} \fr{2\eps
V_d^2}{\eps_n^2-\eps^2} \approx\fr{V_d^2}{\eps}- \eps
\fr{V_d^2}{\eps_B^2}\sum_{n=1}^{n_{\rm max}} \fr{2}{n} \ .
 \ee
The sum over $n$ here diverges only logarithmically and in the
large logarithm approximation it is enough to get only a rough
estimate of the upper cutoff, $n_{\rm max}$. Assuming that the
solutions of the continuous Dirac equation,
Eq.~(\ref{energySingle}), may be used only for energies small
compared to the bandwidth ($\sim t$) gives $\sqrt{n_{\rm
max}}\eps_B\approx t$, or
 \bq\label{nmax}
n_{\rm max}\approx (l/d)^2.
 \ee
The carbon-to-impurity coupling and the carbon-carbon hoping
matrix element are expected to be of the same order of magnitude,
$U\sim t$. This means that in the large logarithm approximation
the left hand side of Eq.~(\ref{KatsMainApproximate}) should be
neglected and the energies of two impurity induced levels become
 \bq\label{levelimpurity}
\eps_{S_\pm}=\pm\fr{\eps_B}{2\sqrt{L}},
 \ee
where $L=\ln(l/d)\gg 1$. In the leading approximation the two
energies $\eps_{S_\pm}$ do not depend on the strength of the
coupling to the impurity, $U$.

The normalization condition Eq.~(\ref{KatsMainMagneticfd}) gives
the probability for the electron to stay at the impurity atom
 \bq
f_{d\pm}^2 \approx\fr{\sqrt{3}\pi}{4} \fr{t^2}{U^2 L}.
 \ee
Since $U/t\sim 1$, the probability to find electron at the adatom
is small like the inverse of a large logarithm.

Now one may found explicitly the impurity induced state wave
function Eq.~(\ref{singularB})
 \bq
\Psi_{S_\pm}=\sum f_n\Psi_n\ ,
 \ee
using the Landau level states $\Psi_n$ coupled to the impurity,
Eq.~(\ref{eigenfunctions}), and the amplitudes $f_n=V_d f_d/(\eps
-\eps_n)$, Eq.~(\ref{KatsMainMagnetic}). In the leading
approximation the energy $\eps$ in the denominator in $f_n\sim
1/(\eps -\eps_n)$ may be neglected for all $n\neq 0$. After that
the contribution to the upper component of $\Psi_{S_\pm}$ from the
Landau level states with $n\neq 0$ vanishes, leading to
 \begin{eqnarray}\label{FindingPoles}
\Psi_{S_\pm} &\sim& \fr{e^{i{\bf
Kr}}}{\eps_{S_\pm}}\left(\begin{array}{cc}
\phi_{0,0} \\
0
\end{array}
\right)\\
&+&\sum_{n=1}^{n_{\rm
max}}\fr{i}{\eps_B\sqrt{n}}\left(\begin{array}{cc}
0 \\
e^{i{\bf Kr}}\phi_{n-1,1} + e^{i{\bf K'r}}\phi_{n,-1}
\end{array}
\right).\nonumber
 \end{eqnarray}
This formula should be compared with the final result for the wave
function, Eq.~(\ref{singularB}). At first sight it seems
surprising: How the lower component of the impurity state
Eq.~(\ref{singularB}) may be singular at small distances, if it is
built from the functions $\phi_{n-1,1}$ and $\phi_{n,-1}$, which
all vanish at $z\rightarrow 0$? To understand this let us consider
the small distance behavior of these functions
 \bq\label{phin-11}
\phi_{n-1,1}\approx \sqrt{\fr{n}{4\pi}} z^*\left(
1-\fr{n|z|^2}{2}+\cdots \right),
 \ee
and
 \bq\label{phin-1}
\phi_{n,-1}\approx -\sqrt{\fr{n}{4\pi}} z\left(
1-\fr{n|z|^2}{2}+\cdots \right).
 \ee
For large $n$ even near the origin both $\phi_{n-1,1}$ and
$\phi_{n,-1}$ are oscillating functions of $|z|$. We keep the
second term of the expansion at small $|z|$ in
Eqs.~(\ref{phin-11},\ref{phin-1}) to show, that the period of this
oscillations scales like $\Delta|z|\sim 1/\sqrt{n}$. Thus the
short distance behavior, for example, of a first sum in
Eq.~(\ref{FindingPoles}) is
 \bq
\sum_n\fr{1}{\sqrt{n}}\phi_{n-1,1}\sim z^*\int_0^{1/|z|^2} dn \sim
\fr{1}{z}\ .
 \ee
This simple estimate does not yet allow to find an overall
numerical factor at the $1/z^*$($1/z$) term. This factor may be
found from the condition of orthogonality of $\Psi_{S_+}$ and
$\Psi_{S_-}$, leading to Eq.~(\ref{singularB}).

More complicated is finding the enveloping function $e^{-|z|^2/4}$
of the lower component of $\Psi_{S_\pm}$ in Eq.~(\ref{singularB}).
Derivation of this long distance behavior is given below.

\subsection{Large distance asymptotics of the singular
state}\label{sec.asymptotic}

First, let us write the spinor wave function
Eq.~(\ref{FindingPoles}) in a form $\Psi_S=e^{i\vec K\vec r}\psi +
e^{i\vec K'\vec r}\psi'$, where the two smooth spinor functions
have a form
 \bq\label{SpinorPolar}
\psi=\left(\begin{array}{cc}
 u(r) \\
 ie^{-i\phi}v(r)
\end{array}
\right), \ \ \psi'=\left(\begin{array}{cc}
 \tilde{u}(r) \\
 ie^{i\phi}\tilde{v}(r)
\end{array}
\right),
 \ee
and we introduced the polar coordinates, $x=r\cos\phi$,
$y=r\sin\phi$. The four functions $u,v,\tilde u,\tilde v$ satisfy
two systems of equations
 \bq\label{system1Mag}
\fr{\eps}{\sqrt{2}\eps_B}v= \fr{du}{dr} +\fr{ 1}{2}r u \ , \
\fr{\eps}{\sqrt{2}\eps_B}u= -\fr{dv}{dr} -\fr{v}{r}+\fr{1}{2}r v,
 \ee
and
 \bq\label{system2Mag}
\fr{\eps}{\sqrt{2}\eps_B}\tilde v= -\fr{d\tilde u}{dr} +\fr{
1}{2}r \tilde u \ , \ \fr{\eps}{\sqrt{2}\eps_B}\tilde u=
\fr{d\tilde v}{dr} +\fr{\tilde v}{r}+\fr{1}{2}r \tilde v.
 \ee
Since the energy, $\eps\ll\eps_B$, is small, one may solve these
equations iteratively: First find the solution for $\eps=0$ and
then look for the corrections $\sim \eps$, $\sim \eps^2$, etc.

For $\eps=0$ Eqs.~(\ref{system1Mag},\ref{system2Mag}) become a set
of decoupled first order differential equations, leading to four
independent (in general non-normalizible) solutions
 \begin{eqnarray}
&&1. \ u=e^{-r^2/4},
v=\tilde u=\tilde v=0, \\
&&2. \ v=\fr{1}{r}e^{r^2/4},
u=\tilde
u=\tilde v=0,\nonumber\\
&&3. \ \tilde u= e^{r^2/4},
v= u=\tilde v=0,\nonumber \\
&&4. \ \tilde v=\fr{1}{r}e^{-r^2/4},
v= u=\tilde
u=0. \nonumber
 \end{eqnarray}
Using only the zero energy solutions which are regular at large
distances, 1. and 4., we find three of the functions introduced in
Eq.~(\ref{SpinorPolar})
 \begin{eqnarray}\label{utildeutildev}
u=e^{-\rho^2/4} \ , \ \tilde u=0 \ , \ \tilde
v=\fr{c}{\rho}e^{-\rho^2/4},
 \end{eqnarray}
where the coefficient $c$ is found from Eq.~(\ref{singularB}).

To find the last function, $v$, we take finite energy $\eps$ and
substitute $u$ Eq.~(\ref{utildeutildev}) into the second equation
(\ref{system1Mag}),
 \begin{eqnarray}
v= \fr{\eps l_B}{\hbar
v_F}\fr{1}{\rho}e^{\rho^2/4}\int_\rho^\infty x e^{-x^2/2}dx
=\fr{\eps l_B}{\hbar v_F}\fr{e^{-\rho^2/4}}{\rho} .
 \end{eqnarray}

\section{Calculating the exchange energy}\label{Appendix_B}

This section describes the derivation of the exchange interaction
matrix elements presented in the main text.

Using Eqs.~(\ref{eigenfunction0}) and (\ref{exchangeMatriD}) the
exchange energy of the electron with the orbital quantum number
$M$ is written in the form
 \begin{align}\label{exchS1}
\eps_{M}=\mp\sum_{m=0}^\infty \int \fr{e^2}{2|\vec r-\vec
r'|}\phi_{0,M}^*(\vec r)\phi_{0,m}(\vec r) \\
\times \phi_{0,m}^*(\vec r')\phi_{0,M}(\vec r')d\vec rd\vec
r'.\nonumber
 \end{align}
The energy is negative for occupied states and positive for empty
ones. Summation over $m$ in Eq.~(\ref{exchS1}) gives the
projection operator onto the occupied $n=0$ Landau Level, which
may be found exactly
 \begin{align}\label{exchS2}
&\rho(\vec r,\vec r')=\sum_{m=0}^\infty
\phi_{0,m}(\vec r)\phi_{0,m}^*(\vec r') \nonumber\\
&= \sum_{m=0}^\infty \fr{(z^* z'/2)^m}{2\pi m!}e^{-|z|^2/4
-|z'|^2/4}\nonumber\\
&= \fr{1}{2\pi}\exp \left\{ -\fr{|z|^2 + |z'|^2-2 z^*
z'}{4}\right\}\nonumber\\
&= \fr{1}{2\pi}\exp \left\{ -\fr{|z-z'|^2}{4} +\fr{ z^*
z'-zz'^*}{4}\right\}.
 \end{align}
Note that the second term in the argument of the exponent in the
last line here is pure imaginary (a phase). The energies are found
from the generating function $I(\alpha)$,
 \bq
\eps_M=\left. \mp\fr{1}{M!}\left(\fr{\partial}{\partial
\alpha}\right)^M I(\alpha)\right|_{\alpha=0},
 \ee
where
 \begin{eqnarray}
I(\alpha)&=&\int\fr{e^2}{2l|z-z'|}\fr{d^2z d^2z'}{4\pi^2}
\nonumber\\
&\times&\exp \left\{ -\fr{|z|^2 + |z'|^2- z^* z' -\alpha z
z'^*}{2}\right\}.
 \end{eqnarray}
Introducing new variables $u=z-z'$, $v=z+z'$ one finds
 \bq
I(\alpha)= \fr{e^2\sqrt{\pi}}{2\sqrt{2}(1-\alpha)l},
 \ee
leading to Eq.~(\ref{exchangeBareL}).

Exchange dominated Zeeman splitting of the singular impurity level
Eq.~(\ref{singularB}) is calculated with the help of the density
operator Eq.~(\ref{exchangeMatrix}). The exchange energy in this
case is dominated by the interaction of the impurity level with
itself,
 \bq\label{exchangeSsup}
\eps_S=\mp2\int \fr{e^2}{|\vec r - \vec r'|} |\Psi_{S_1}(\vec
r)|^2 |\Psi_{S_2}(\vec r')|^2 d\vec r d\vec r'.
 \ee
Here $\Psi_{S_1}$ and $\Psi_{S_2}$ are the upper and lower
components of the spinor function $\Psi_S$ Eq.~(\ref{singularB}).
The energy Eq.~(\ref{exchangeSsup}) formally coincides with the
Coulomb interaction energy of two charge densities
$|\Psi_{S_1}(\vec r)|^2$ and  $|\Psi_{S_2}(\vec r')|^2$. The
calculation of the energy is greatly simplified after one notices
that the density $|\Psi_{S_2}(\vec r')|^2$ is mostly concentrated
at distances small compared to the Larmor radius, $r\ll l$, and
the total charge in each component of the wavefunction is $1/2$.
Thus for $\ln(l/d)\gg 1$
 \bq\label{exchangeSsupF}
\eps_S=\mp \int \fr{e^2}{r} |\Psi_{S_1}(\vec r)|^2 d\vec r = \mp
\fr{e^2}{2l}\sqrt{\fr{\pi}{2}}.
 \ee

The exchange dominated splitting of $n=0$ Landau level,
Eq.~(\ref{exchangeBareL}), does not depend on the angular quantum
number $M$, as is it should be for the translationally invariant
system. The $M$-dependance appears in the case of impurity adatom
considered in this paper. As was discussed in the main text of the
paper, the electron states with $M\neq 0$ from the $n=0$ Landau
level effectively know about the impurity because of missing the
exchange interaction with the $n=0, m=0$ electron. Corresponding
correction to the energy is given by the same formula
Eq.~(\ref{exchS1}), where one leaves only the $m=0$ term in the
sum, i.e.
 \begin{align}
 &\Delta\eps_M=\left. \pm\fr{1}{M!}\left(\fr{\partial}{\partial
\alpha}\right)^M J(\alpha)\right|_{\alpha=0},\nonumber\\
 &J(\alpha) = \int \fr{e^2}{2l|z-z'|} \fr{d^2z
d^2z'}{4\pi^2}\\
&\times\exp \left\{ -\fr{|z|^2 + |z'|^2 -\alpha z z'^*}{2}\right\}
=\fr{e^2}{4l}\sqrt{\fr{2\pi}{2-\alpha}}.\nonumber
 \end{align}
This leads to Eq.~(\ref{exchangeSm}).

\section{Solving the Dirac equation with several impurities}\label{Appendix_C}

In order to find the energies of impurity induced states from
Eq.~(\ref{GabN0}) one needs to know the matrix $G_{ab}$,
describing the electron's hoping between the impurity sites. In
case of an atom placed at the origin, $z_a=(x_a+iy_a)/l=0$, and
for the vector potential in a polar gauge, $\vec A=({By}/{2},
-{Bx}/{2},0)$, the matrix element $G_{aa}$ was already found in
Eq.~(\ref{KatsMainApproximate})
 \bq\label{GaaN}
G_{aa}= \fr{{V_d^2}}{\eps}\left[1 - 4\fr{\eps^2}{\eps_B^2}\ln
\fr{l_B}{d}\right].
 \ee
We will see in a moment, that this formula works for any diagonal
element of the matrix $G_{ab}$.

Eq.~(\ref{GaaN}) is valid in the limit $\ln(l/d)\gg 1$. To reach
the better accuracy one would need to go beyond the Dirac equation
approximation and to find the electrons wavefunctions on the
hexagonal lattice in magnetic field. Consequently, corrections of
higher orders in small energy, $\sim\eps^3$, are also neglected in
Eq.~(\ref{GaaN}). (The same holds for the accuracy of
Eqs.~(\ref{GabN1}) and (\ref{GabN2}) below.)

Suppose now that the atom $b$ is placed at the origin ($z_b=0$)
and the atom $a$ is not. Calculation of the element of the matrix
$G_{ab}$ now formally coincides with the calculation of the upper
component of the singular impurity state wavefunction
Eq.~(\ref{FindingPoles}), which gives
 \bq\label{GabN1}
G_{ab}= \fr{V_d^2}{\eps}\left[1 - 4\fr{\eps^2}{\eps_B^2}\ln
\fr{1}{|z_{ab}|}\right] e^{-{|z_{ab}|^2}/{4}}.
 \ee
Here $z_{ab}=z_a -z_b$ and $|z_{ab}|=r_{ab}/l$, where $r_{ab}$ is
the true distance between the atoms $a$ and $b$.

The value of the matrix element $G_{ab}$, Eq.~(\ref{GabN1}),
depends only on the distance between two atoms, $|z_{ab}|$. Thus
one is tempted to use this formula in case of arbitrary positions
of both atoms. However, for the equation~(\ref{GabN1}) to be valid
it is also important that the vector potential, $\vec A$, also
vanishes at the position of one of the atoms $a$ or $b$. The shift
of the position of vanishing of (the both components of) the
vector potential is achieved by the simple gauge transformation,
adding a factor
 \bq\label{phaseN}
\exp\left\{i\fr{x_ay_b-y_ax_b}{2l^2}\right\}
 \ee
to the equation~(\ref{GabN1}). After taking into account this
phase the formula for the element of the matrix may be written in
a simple form
 \begin{eqnarray}\label{GabN2}
G_{ab}&=& \fr{V_d^2}{\eps}\left[1 - 4\fr{\eps^2}{\eps_B^2}\ln
\fr{1}{|z_{ab}|}\right] \\
&\times&\exp\left\{ -\fr{|z_a|^2+|z_b|^2-2z_a z_b^*}{4}
\right\},\nonumber
 \end{eqnarray}
where now the "coordinates" $z_a$ and $z_b$ are measured from the
point of vanishing of the vector potential, which is chosen to be
somewhere in the middle of the bunch of impurities.

It is convenient to introduce the new variables $\phi_a$,
 \bq\label{ftophi}
\phi_a =e^{-{|z_a|^2}/{4}} f_a,
 \ee
and rewrite Eq.~(\ref{GabN0}) in a form
 \begin{eqnarray}
&&(\eps -\eps_a) e^{|z_a|^2/2}\phi_a
=\\
&&=\fr{V_d^2}{\eps}\sum_b\left[1 - 4\fr{\eps^2}{\eps_B^2}\ln
\fr{1}{|z_{ab}|}\right] \exp\left\{ \fr{z_a^*
z_b}{2}\right\}\phi_b.\nonumber
 \end{eqnarray}
In the large logarithm approximation since $\eps_a\ll \eps_B$ and
$V_d\sim\eps_B$, the left hand side of this equation should be
omitted, leading to
 \bq\label{logN}
\sum_b\exp\left\{ \fr{z_a^* z_b}{2}\right\}\phi_b
=4\fr{\eps^2}{\eps_B^2}\sum_b \ln \fr{1}{|z_{ab}|} \exp\left\{
\fr{z_a^* z_b}{2}\right\}\phi_b.
 \ee
This is the final equation, written in the most compact form,
which one needs to solve in order to find the energies and wave
functions of the impurity induced localized states. There is
certain freedom in the choice of positions of the origin, $z=0$,
which may be (but don't necessary have to) fixed by e.g. requiring
vanishing of the average distance $\sum_a z_a=0$.

Equation~(\ref{logN}) is an eigenvalue problem of the form $A\phi
=\eps^2 B\phi$, where $A$ and $B$ are two Hermitean matrices. What
makes this problem tractable is that in case of all $|z_a|\ll 1$,
the matrix in the left hand side of the equation, $A_{ab}=
e^{z_a^*z_b/2}$, has eigenvalues of a very different magnitude.
Indeed, one may expand the exponent $e^{z_a^*z_b/2}$ in a power
series and found (for a bunch of impurities centered as $\sum
z_a=0$) the three first largest eigenvalue $N$,
$\fr{1}{2}\sum|z_a|^2$, and $\fr{1}{8}[\sum|z_a|^4-\fr{1}{N}|\sum
z_a^2|^2 -|\sum z_a|z_a|^2|^2/(\sum |z_a|^2)]$. Each next
eigenvalue (having more and more complicated explicit form) will
contain higher powers of small $|z_a|^2$.

We will show now that this property of the matrix $e^{z_a^*z_b/2}$
leads to a hierarchical structure of the eigenvalues of
Eq.~(\ref{GabN0})
 \bq
|\eps^{(1)}|\gg |\eps^{(2)}|\gg |\eps^{(3)}| \cdots .
 \ee
We use superscript indices to enumerate the energies,
$\eps^{(i)}$, to avoid confusion with the Landau level and
impurity atom energies, $\eps_n$, $\eps_a$. As always, in the
particle-hole symmetric limit, $\eps_a\rightarrow 0$, the
eigenvalues appear in $\pm$ pairs, as is obvious from the
Eq.~(\ref{logN}).

To investigate the properties of the solutions of Eq.~(\ref{logN})
let us consider the series of several consecutive approximations
to this equation.

{\em First iteration.} At first step we neglect completely all the
small arguments in the exponent $e^{z_a^*z_b/2}$ both in the left
and right hand sides of Eq.~(\ref{logN}), leading to
 \bq\label{logN0}
\sum_b\phi_b =4\fr{\eps^2}{\eps_B^2}\sum_b \ln \fr{1}{|z_{ab}|}
\phi_b.
 \ee
Remarkably, this equation has $N-1$ exact zero energy solutions,
$\phi_a^{(i)}$, satisfying a single constraint
 \bq\label{constraint1}
\sum_b \phi_b^{(i)}=0, \ \eps^{(i)}\equiv 0.
 \ee
The last and only nontrivial solution in the limit of all large
logarithms has the energy
 \bq\label{LargeLogEnN}
\eps\approx \pm\fr{\eps_B}{2\sqrt{\ln(1/\langle
|z_{ab}|\rangle)}},
 \ee
where $\langle |z_{ab}|\rangle$ is some average distance between
impurities. Corresponding eigenfunction has a simple form,
 \bq\label{LargeLogPhiN}
\phi_a\approx 1/\sqrt{N},
 \ee
only if not only all the logarithms in the right hand side of
Eq.~(\ref{logN0}) are large, but also if there are only two
different large logarithms, $L_1=\ln (a/d)$ (in case $a=b$) and
$L_2=\ln(1/|z_{ab}|)$ (in case $a\neq b$). This means that the
distance between any two impurities in the bunch is of the same
(close) order of magnitude.

Each amplitude $\phi_a$ via Eqs.~(\ref{ftophi}) and (\ref{fnN})
generates a contribution of the form Eq.~(\ref{singularB}) to the
electrons wave function in graphene, centered at the impurity $a$.
Even though Eqs.~(\ref{LargeLogEnN}) and (\ref{LargeLogPhiN}) have
a rather poor (at best $\sim 1/\log$) accuracy, they are enough to
make strong conclusions about the wave function of the largest
energy impurity state. The upper component of the wave function
consists of the sum of many $m=0$ functions with slightly offset
centers and slightly different phases (like in
Eq.~(\ref{phaseN})), which however due to $|z_a|\ll 1$ form an
almost unperturbed joint $m=0$ state. The lower component consists
of many singular $\sim 1/z$ states~\cite{Pereira06} centered at
individual adatoms in the lower component. Exactly as it was in
the case of single adatom, due to the orthogonality of '$+$' and
'$-$' states, Eq.~(\ref{LargeLogEnN}), both of them should have
equal probability to find electron in upper and lower component.
Outside the compact bunch of impurities the electron wave function
for the highest energy impurity state has the form (compare to
Eq.~(\ref{singularB}))
 \bq\label{singularBN0}
\Psi_{m=0} =
\left(\begin{array}{cc}
 e^{i{\bf Kr}}{\displaystyle\fr{e^{-z z^*/4}}{2\sqrt{\pi}}}
 \\
 \pm
{\displaystyle(\alpha \fr{e^{i{\bf Kr}}}{z}+\beta \fr{e^{i{\bf
K'r}}}{z^*})} {\displaystyle \fr{e^{-z z^*/4}}{2\sqrt{\pi L}}}
\end{array}
\right).
 \ee
Here $\alpha$ and $\beta$ are two complex numbers, $|\alpha|^2
+|\beta|^2=1$, and we remind that $\vec K' =-\vec K$. Also the
large logarithm here is $L=\ln (1/\langle |z_{ab}|\rangle)\gg 1$,
where $\langle |z_{ab}|\rangle \ll 1$ is the typical distance
between impurities.

Even more interesting are the other $N-1$ low energy states
described by Eq.~(\ref{constraint1}). The constraint $\sum_b
\phi_b^{(i)}=0$ means that the amplitude of $n=0, m=0$ state in
their upper component (almost) vanish. Thus these states are the
superpositions of {\it other than} $m=0$ states in the upper
component (i.e. $m=1,2,\cdots ,N-1$).

{\em Second iteration.} Our next step will be to expand the
exponent $e^{z_a^*z_b/2}$ in the left hand side of
Eq.~(\ref{logN}) to the first order in a small argument
$z_a^*z_b/2$, leading to
 \bq\label{logN1}
\sum_b\phi_b + \fr{z_a^*}{2} \sum_b
z_b\phi_b=4\fr{\eps^2}{\eps_B^2}\sum_b \ln \fr{1}{|z_{ab}|}
\phi_b.
 \ee
Keeping the same, $\sim z_a^*z_b$, terms in the right hand side of
Eq.~(\ref{logN}) would not add any new qualitative features to the
solution.

Instead of Eq.~(\ref{constraint1}) now any vector $\phi_a^{(i)}$,
satisfying two simple constraints would be an exact zero energy
solution of the system of equations~(\ref{logN1})
 \bq
\sum_b \phi_b^{(i)}=0, \ \sum_b z_b \phi_b^{(i)}=0,\
\eps^{(i)}\equiv 0.
 \ee
Thus there are $N-2$ exact zero energy solutions.

Among the remaining two nontrivial solutions one, with the larger
energy, was described by
Eqs.~(\ref{LargeLogEnN},\ref{LargeLogPhiN}). Second solution for a
center of the bunch chosen to satisfy $\sum z_a=0$ has the form
 \bq
\phi_a\sim z_a^* \ , \ \eps\approx \pm\eps_B\sqrt{\fr{\sum
|z_b|^2}{8(L_1-L_2)}} \ ,
 \ee
where $L_1$ and $L_2$ are two large logarithms defined below
Eq.~(\ref{LargeLogPhiN}).

One may continue expanding the matrix exponent $e^{z_a^* z_b/2}$
in the left hand side of Eq.~(\ref{logN}) to higher orders in the
small argument, to find the finite values of smaller and smaller
eigenvalues. The resulting estimate of the energy values is given
by Eq.~(\ref{Hierarchy}). The wave function for the $m$-th state
outside the small bunch of impurities is given by
Eq.~(\ref{singularBN}). As we told, the in-plane wave function for
many impurities is build as a sum of single-impurity solutions
Eq.~(\ref{singularB}), centered at individual impurities and with
a gauge factor accounting for the center displacement. Since for
$m>0$ these single-impurity contributions strongly cancel each
other, there is no $\sim e^{i\vec K \vec r}/z$ contribution in the
lower component of Eq.~(\ref{singularB}). Contributions $\sim
e^{i\vec K' \vec r}/z^*$ in the lower component of the
pseudospinor come with the phase which is not synchronized with
the phase of the the upper component and thus are not suppressed.
The upper components of different impurity states each acquire an
(almost) well defined and different value of the angular momentum
$m$.

\section{Exchange interaction in case of many adatoms}\label{Appendix_D}

Since the lower component of all states Eq.~(\ref{singularBN}) is
similar, but their upper components have different angular
behavior, it follows from Eq.~(\ref{exchangeMatrix}) that these
states may interact via exchange only with themselves. Calculation
of the exchange energy essentially repeats that performed in
Eqs.~(\ref{exchangeSsup}, \ref{exchangeSsupF}). In the final
result Eq.~(\ref{exchangeSsupF}) one simply need to replace
$\Psi_{S_1}$ by $\phi_{0m}/\sqrt{2}$,
 \bq\label{exchangeSsupFN}
\eps_S^{(m)}=\mp \fr{1}{2}\int \fr{e^2}{r}|\phi_{0m}|^2 d^2r
=\pm\fr{(2m)!}{(m!)^2 2^{2m}}\sqrt{\fr{\pi}{2}}\fr{e^2}{2l}.
 \ee
This formula includes Eq.~(\ref{exchangeSsupF}) as the $m=0$ case.

The exchange energy for the electrons outside the unpolarized
droplet is given by the modified Eq.~(\ref{exchS1})
 \begin{align}\label{exchS1M}
\eps_{M}=\mp\sum_{m=N}^\infty \int \fr{e^2}{2|\vec r-\vec
r'|}\phi_{0,M}^*(\vec r)\phi_{0,m}(\vec r) \\
\times \phi_{0,m}^*(\vec r')\phi_{0,M}(\vec r')d\vec rd\vec
r',\nonumber
 \end{align}
where the summation over $m$ excludes the electrons from the $n=0$
Landau level from inside the large droplet, $N\gg 1$. This
exclusion of states with $m<N$ leads to the projection operator
 \begin{align}\label{exchSN}
\rho_N(r,r')&=\sum_{m=N}^\infty \phi_{0,m}(r)
\phi_{0,m}^*(r')=\nonumber\\ &= \sum_{m=N}^\infty \fr{(z^*
z'/2)^m}{2\pi m!}e^{-|z|^2/4 -|z'|^2/4}.
 \end{align}
Since the function $|\phi_{0,m}(r)|$ has a very pronounced maximum
at $r\approx \sqrt{2m}$, the modified projection operator
Eq.~(\ref{exchSN}) coincides with Eq.~(\ref{exchS2}) in case of
both arguments larger than the droplet size, $r,r'>\sqrt{2N}$, and
vanishes fast if one argument falls inside the droplet. The simple
calculation helps to quantify this observation. Consider a complex
variable $v$ with ${\rm Re}\, v>0$. For $N\gg 1$ one finds
 \begin{align}
e^{-v}\sum_{m=N}^\infty \fr{v^m}{m!}
\approx \sum_{m=N}^\infty
\fr{e^{-(v-m)^2/2m}}{\sqrt{2m\pi}} 
\approx \fr{1}{2}\, {\rm erfc}\left( \fr{N-v}{2\sqrt{N}} \right),
 \end{align}
where a complimentary error function ${\rm erfc}(z)=
(2/\sqrt{\pi})\int_z^\infty e^{-t^2} dt$. Combining this with
Eq.~(\ref{exchS2}) gives
 \begin{align}
\rho_N(r,r')=&\fr{1}{2\pi}\exp\left\{ -\fr{|z-z'|^2}{4}+ i{\rm
Im}\fr{z^* z'}{2}\right\}\times\nonumber\\
&\times \fr{1}{2}{\rm erfc}\left(\fr{2N-z^*z'}{4\sqrt{N}}\right).
 \end{align}
The first exponential factor here shows that the exchange
interaction is effective only at distances of the order of Larmor
radius, $|\vec r -\vec r'|\sim 1$. The second factor, the error
function, guaranties vanishing of the interaction inside the
droplet, at $|\vec r|<\sqrt{2N}$.

After calculating the sum over $m$ in Eq.~(\ref{exchS1M}) one
still needs to perform the integration over two coordinates $\vec
r, \vec r'$. We don't see an easy way to perform these
integrations in a compact form. However, the resulting behavior of
the energy $\eps_M$ is clear. For large angular momentum electron
stays at the circle larger than the droplet radius and has the
same energy as the electrons at the $n=0$ Landau level have in
clean graphene without impurities, Eq.~(\ref{exchangeBareL}).
Exactly at the border of the droplet the exchange dominated Zeeman
splitting drops by half
 \bq\label{ZeedrN}
\eps_{M\gg N}= \mp\fr{e^2}{2l}\sqrt{\fr{\pi}{2}} \ , \ \eps_{M=N}=
\fr{1}{2}\eps_{M\gg N} \ .
 \ee
For $M$ slightly bigger than $N$ the Zeeman splitting increases
continuously between the two values, reaching the asymptotic value
at $M-N\gg\sqrt{N}$.

\end{document}